# Implementing a simple vectorial bridge with a digital oscilloscope

Rosario Bartiromo[1]

Istituto di Struttura della Materia del CNR, via Fosso del Cavaliere 100, 00133 Roma, Italy
and
Dipartimento di Fisica, Università di Roma Tre, via della Vasca Navale 84, 00146 Roma, Italy

Mario De Vincenzi

Dipartimento di Fisica, Università di Roma Tre, via della Vasca Navale 84, 00146 Roma, Italy
and
Istituto Nazionale di Fisica Nucleare, Sezione di Roma3, via della Vasca Navale 84, 00146 Roma, Italy

**Abstract**

We show how to exploit instrumentation available in undergraduate student laboratories to build a simple vectorial bridge. In particular we take advantage of the possibility to read data from a digital oscilloscope with a personal computer and we describe an algorithm to obtain an accurate evaluation of the phase difference between two sinusoidal signals. The use of the bridge to characterize components of a high Q RLC filter is shown to greatly improve the understanding of results in resonance experiments. Direct evidence of dielectric losses, of skin currents and of the effect of distributed capacitance is obtained.

---

[1] e-mail: bartiromo@fis.uniroma3.it

I  Introduction

The study of resonant circuits, both in theory and experiments, is an important step in the learning path of physics students. RLC systems provide a paradigmatic tool to introduce young learners to the concept of normal modes and to the methodologies of linear system analysis.[1,2,3] In the laboratory RLC circuits offer the opportunity to familiarize with instruments of fundamental importance in the future professional life such as oscilloscopes, waveform generators and data acquisition with personal computers.[4]

It is therefore rather disturbing that it is never easy to get a good matching when it comes to comparing experimental results obtained with high Q RLC circuits with theoretical predictions[2,5]. These difficulties stem from shortcomings in the modelling of real components and specifically from distributed capacitance and frequency dependent dissipation caused by skin effect, dielectric or magnetic losses. These effects are very difficult to compute from theory thus making a quantitative evaluation from first principles impossible. A viable alternative consists in a better characterization of the available components obtained by measuring their complex impedance as a function of the excitation frequency. This can be done with a variable frequency vectorial bridge[6] which however is not often available in student laboratories.

In this paper we show that the same instrumentation used for the RLC experiment can be exploited to build such an instrument. In particular we will take advantage of the possibility to read the data from a digital oscilloscope with a personal computer and we will describe an algorithm to obtain a very accurate evaluation of the phase difference between two sinusoidal signals.

Indeed any method aiming at quantifying dissipation in reactive components boils down to measuring the phase angle $\phi$ between the voltage applied and the current flowing in the component. The relative accuracy achievable on the equivalent resistance $R_X$ turns out to be linked to the phase measurement accuracy $\delta\phi$ by the following relation $\dfrac{\delta R_X}{R_X} \geq \dfrac{X(\omega)}{R_X}\delta\phi$. Since the reactance $X(\omega)$ can be a couple of orders of magnitude larger than $R_X$, an accuracy of 10% for the dissipation measurement requires phase accuracy of the order of $10^{-3}$ radiant, i. e. 0.05°. We will show that this accuracy is achievable with a careful exploitation of instruments of current use in a student laboratory.

The structure of the paper is the following: phase measurements with a digital oscilloscope and their uncertainties are described in the next two sections; a very simple bridge configuration is then presented in section IV together with the expressions to evaluate both resistance and reactance with their uncertainties. Then in section V we illustrate the difficulties that may be encountered in accounting for experimental observations with a high Q resonant RLC filter by discussing a couple of exempla. Finally in section VI we show the results obtained for the frequency dependence of the impedance of the capacitors and the inductor used in the measurements. The last section presents the conclusions of our work.

II  Phase measurements with a digital oscilloscope

In the realm of analogical electronics we could obtain the phase difference $\phi$ of two signals $V_{in}(t) = V_{in}\sin(\omega t)$ and $V_{out}(t) = V_{out}\sin(\omega t + \phi)$ by using a device that performs their product. We would then deal with a signal

$$V(t) = kV_{in}V_{out}\sin(\omega t)\sin(\omega t + \phi) = \frac{kV_{in}V_{out}}{2}[\cos(\phi) - \cos(2\omega t + \phi)]$$

and we could measure the alternate and the continuous component to recover the cosine of the phase angle from the value of their ratio. If we measure V(t) with an oscilloscope then it would be sufficient to measure its maximum M and minimum m value to obtain $\cos(\phi)$ = (M+m)/(M-m).

With a double channel digital oscilloscope we can obtain a better accuracy by measuring directly the two original voltage signals and reading the data files with a personal computer. Indeed a number of methods can be found in the published literature aiming at optimally recovering the phase difference between two digitized sinusoidal signals[7] and recommended standards have been issued by the IEEE.[8,9] In the following we will describe in some detail an algorithm that:
   i. is best suited for use in an impedance meter,
   ii. takes into account all relevant features of an analogical to digital converter,
   iii. allows for a clear evaluation of relevant uncertainties,
   iv. can be usefully illustrated in a single undergraduate classroom session.

Neglecting noise for the time being, we will have to deal with a digital representation for each signal consisting of a number $N_p$ of data points taken at a constant and precisely known time interval $\Delta t$:

$$V_{in}(t_i) = K_{in} V_{in} \sin(\omega t_i) + \delta_i^{in} + \delta_0^{in} \quad \text{and} \quad V_{out}(t_i) = K_{out} V_{out} \sin(\omega t_i + \phi) + \delta_i^{out} + \delta_0^{out}$$

where the coefficients K represent the ADC's overall calibration factor, $\delta_0$ their offset value and $\delta_i$ are the quantization errors.

Taking their product we see that a number of additional terms interfere with the phase information we are looking for and we must properly handle them to minimize their impact:

$$V_{in}(t_i)V_{out}(t_i) = K_{in}K_{out}V_{in}V_{out}\sin(\omega t_i)\sin(\omega t_i + \phi) + \\ K_{in}V_{in}\sin(\omega t_i)(\delta_i^{out} + \delta_0^{out}) + K_{out}V_{out}\sin(\omega t_i + \phi)(\delta_i^{in} + \delta_0^{in}) + (\delta_i^{out} + \delta_0^{out})(\delta_i^{in} + \delta_0^{in}) \quad (1)$$

Since quantization errors are in principle not correlated, we are lead to work with averages of this expression. Therefore, assuming the knowledge of the signal period and deferring a discussion of the impact of its uncertainty to the following section, we will average over an integer number of periods so that the second and the third term have a null expected value. Then its first term yields a numerical evaluation of $K_{in}K_{out}V_{in}V_{out}\cos(\phi)/2$ whose accuracy we will comment later.

For the last term we note that since the two factors are not correlated, its average can be recovered from the product of their averages. Using the symbol $<>$ to denote our numerical averages we have

$$<V_{in}(t_i)> = <\delta_i^{in} + \delta_0^{in}> \quad \text{and} \quad <V_{out}(t_i)> = <\delta_i^{out} + \delta_0^{out}>$$ from which we finally get the average of expression (1) as

$$<V_{in}(t_i)V_{out}(t_i)> = K_{in}K_{out}V_{in}V_{out}\cos(\phi)/2 + <V_{in}(t_i)><V_{out}(t_i)>$$

The values of $V_{in}$ and $V_{out}$ can be recovered numerically by evaluating the root mean square of the two signals. Proceeding as above we get for the input signal

$$V_{in}(t_i)^2 = K_{in}^2 V_{in}^2 \sin(\omega t_i)^2 + 2K_{in}V_{in}(\delta_i^{in} + \delta_0^{in})\sin(\omega t_i) + (\delta_i^{in} + \delta_0^{in})^2 \quad (2)$$

and averaging this expression we obtain

$$K_{in}^2 V_{in}^2 / 2 = <V_{in}(t_i)^2> - <V_{in}(t_i)>^2 - \sigma_{in}^2$$

where $\sigma_{in}^2 = <(\delta_i^{in} + \delta_0^{in})^2> - <(\delta_i^{in} + \delta_0^{in})>^2$ is the variance of the distribution of quantization errors.

We can write similar expression for the output channel and finally get the phase cosine as

$$\cos(\phi) = \frac{<V_{in}(t_i)V_{out}(t_i)> - <V_{in}(t_i)><V_{out}(t_i)>}{\sqrt{<V_{in}(t_i)^2> - <V_{in}(t_i)>^2 - \sigma_{in}^2}\sqrt{<V_{out}(t_i)^2> - <V_{out}(t_i)>^2 - \sigma_{out}^2}} \quad (3)$$

that does not depend upon the calibration factors $K_{in}$ and $K_{out}$ whose uncertainties therefore do not affect this measurement.

Only the phase absolute value can be recovered from the knowledge of its cosine but the availability of digital data allows determining its sign too. When the frequency is known, we can add a digital phase delay to the output signal by shifting it with respect to the input signal of a number of data point corresponding to a quarter of the period, that is a phase shift of $\pi/2$, and use the same algorithm to obtain $\cos(\phi+\pi/2)=\sin(\phi)$ from which the sign of the phase angle is deduced.

Using a Taylor expansion, it can be shown that the numerical evaluation of the averages involved in our phase algorithm does not produce a significant error when the average is performed on an integer number of cycles. This is of course not possible when the ratio of the signal period to the sampling time is not rational and results in a phase uncertainty that can be easily shown to be lower than 1/N, N being the number of data points used in the averaging process. In the following we will take care to use N > 2000 to make sure that this uncertainty stays below our target of $\delta\phi < 10^{-3}$ radiant.

We have used a series of experimental measurements to compare the phase obtained with eq. (3) with the results of a non linear fit of the measured signals performed by MINUIT, a collection of minimization libraries developed at CERN.[10] As shown in table 1 the agreement is excellent: the difference between the two values always stays below the combined statistical uncertainty and on average does not exceed one part over a thousand. This comparison fully validates the algorithm discussed in this section.

| phase (fit) | phase (eq.3) |
|---|---|
| 0.3±0.1° | 0.42 ±0.05° |
| 0.37±0.01° | 0.37 ±0.02° |
| 0.39±0.01° | 0.39 ±0.02° |
| 0.45±0.01° | 0.44 ±0.02° |
| 27.34±0.01° | 27.34 ±0.02° |
| 28.05±0.01° | 28.06 ±0.02° |
| 78.54±0.01° | 78.55 ±0.02° |
| 86.1±0.1° | 86.09 ±0.05° |

Table I

III Phase uncertainties

Statistical uncertainties in the phase measurements are caused by fluctuations in the value of the quantization errors. Starting from expression (1), (2) and (3), a long but straightforward calculation, see appendix, shows that to the leading order in the size of quantization errors we have

$$\sigma^2_{\cos(\phi)} = \frac{\sin^2(\phi)}{N}\left(\frac{\sigma_{in}^2}{<V_{in}^2(t_i)>} + \frac{\sigma_{out}^2}{<V_{out}^2(t_i)>}\right) \quad (4)$$

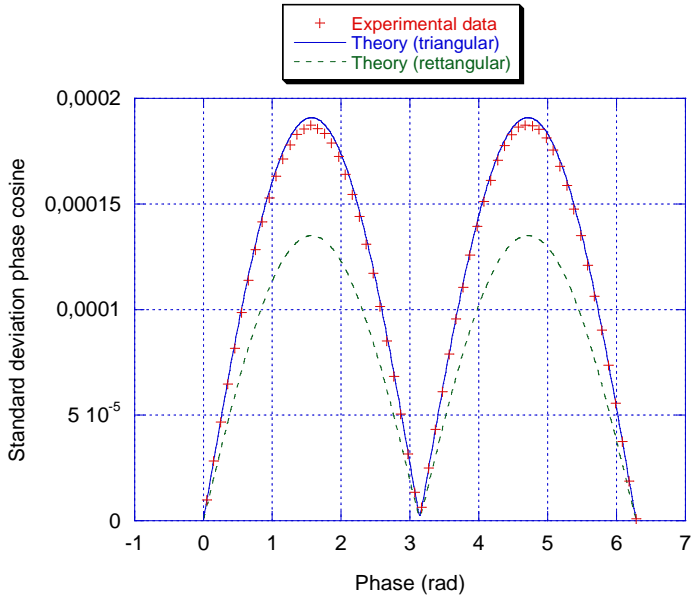

Fig. 1: Standard deviation of cos(φ) distribution obtained from a series of 10 identical measurement of the phase difference between input and output signal across a 9/10 voltage divider. Input amplitude 5Vrms, signal frequency 800 Hz, 2500 data points. The two signals have been digitally shifted to study the phase dependence of the measurement accuracy.

The two quantities $\sigma_{in}^2$ and $\sigma_{out}^2$ depend on the distribution of ADC's quantization errors. Two alternative results are reported in the literature. Denoting the voltage equivalent of the bit as $\delta V_b$, the most popular assumption[11] is that quantization errors are uniformly distributed between $-\delta V_b/2$ and $+\delta V_b/2$: in this case their variance is $\frac{\delta V_b^2}{12}$. More recently it has been argued[12] that this distribution is triangular and symmetric between $-\delta V_b$ and $+\delta V_b$ yielding a variance equal to $\frac{\delta V_b^2}{6}$.

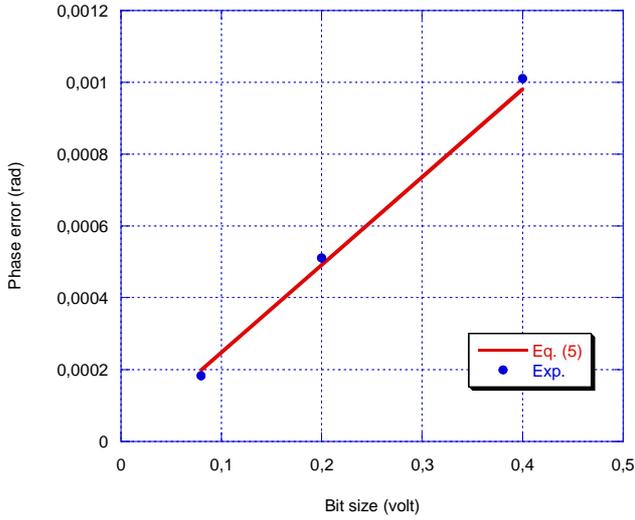

*Fig. 2: Standard deviation of ϕ distribution obtained from a series of 10 identical measurement of the phase difference between input and output signal across a 9/10 voltage divider as a function of the equivalent bit size. Input amplitude 5Vrms, signal frequency 800 Hz, 2500 data points.*

To identify the correct assumption between these two options, we have conducted a series of repeated measurements of the input and output signals of a 9/10 resistive divider with a Tektronics TDS1012 two channel digital storage oscilloscope used in single sweep to avoid fluctuations from trigger jitters. The phase shift dependence of the standard deviation of the results has been obtained with the technique of the digital phase delay. Experimental data are compared in fig. 1 with predictions of equation (4) and show a very good agreement with the assumption of a triangular distribution of quantization errors. This allows the use of expression (4) to quantify statistical phase uncertainty in the impedance measurements described in the following of this paper.

This finding implies that quantization errors cause a statistical fluctuation δϕ of the phase that does not depend upon the phase value and is determined only by the ratio of the bit size to the signal amplitude. We have tested this result in fig. 2 showing that the measured phase fluctuation for fixed signal amplitude is well described by eq. (4) when the vertical range of the oscilloscope channels is increased to change the bit size. This figure also shows that adapting the vertical range of the oscilloscope channels to the signal amplitudes we can obtain a statistical uncertainty on the phase angle small enough to be useful in impedance measurements.

The impact of electrical noise on phase accuracy can be evaluated proceeding as in the previous section. It is easy to demonstrate that, since noise is not correlated with quantization errors, we have:

$$\sigma_\phi^2 = \frac{1}{N}\left[\frac{\sigma_{in}^2 + <(\delta V_{noise}^{in})^2>}{<V_{in}^2(t_i)>} + \frac{\sigma_{out}^2 + <(\delta V_{noise}^{out})^2>}{<V_{out}^2(t_i)>}\right] \qquad (5)$$

We have again tested successfully this result by changing the signal amplitude in experiments with a fixed noise level.

Noise is seldom important in the input channel but it can significantly affect the output signal when it becomes too small, depending on the impedance we are attempting to measure. It can be shown using eq. (5) that, to be compatible with our accuracy goal, the signal to noise ratio in the output channel should not fall below the value of $\frac{10^3}{\sqrt{N}} \approx 25$ for N=2000.

It is important to stress at this point that the presence of noise also affects the phase evaluation. With the method used above, and assuming that noise in the input and output channel are not correlated, it is easily shown that the previous expression (3) must be modified as

$$\cos(\phi) = \frac{<V_{in}(t_i)V_{out}(t_i)> - <V_{in}(t_i)><V_{out}(t_i)>}{\sqrt{<V_{in}(t_i)^2> - <V_{in}(t_i)>^2 - \sigma_{in}^2 - <(\delta V_{noise}^{in})^2>}\sqrt{<V_{out}(t_i)^2> - <V_{out}(t_i)>^2 - \sigma_{out}^2 - <(\delta V_{noise}^{out})^2>}}$$

To the leading order the noise amplitude introduces a systematic overestimate of the phase cosine equal to a fraction $\frac{1}{2}\frac{<(\delta V_{noise}^{out})^2>}{<V_{out}^2>}$ of its value. With the limit stated above on the signal to noise ratio we can neglect the noise bias except when measuring small phase angle where it can translate into an important phase error. To circumvent this problem, when the phase angle is below 30° we introduce a digital phase delay of 45°, evaluate the phase difference between the two signals and then subtract the digital delay to obtain a better evaluation of the original phase angle. This allows making negligible the noise bias with respect to the statistical uncertainty also in these unfavourable circumstances.

Before closing this section let us note for future use that the statistical uncertainties on the effective input signal amplitudes can be computed starting from eq. (2) by the method outlined in the appendix and is given by $\frac{\sigma_{V_{in}}}{V_{in}} = \frac{1}{\sqrt{N}}\left[\frac{\sigma_{in}^2 + <(\delta V_{noise}^{in})^2>}{<V_{in}^2(t_i)>}\right]^{1/2}$ with a similar expression for the output channel.

## IV A simple vectorial bridge

The simplest experimental arrangement for impedance measurements that allows exploiting the opportunities offered by a digital oscilloscope is illustrated in fig. 3. We build a voltage divider by connecting the unknown impedance Z in series with a reference impedance $Z_0$. An Agilent 33120A function generator is used to power the circuit while two well compensated 1/10 probes (Tektronics TPP0101) are used to pick up the input and the output signals that are sampled by the two measuring channels of a Tektronics TDS1012 digital storage oscilloscope whose memory buffer is read by a personal computer.

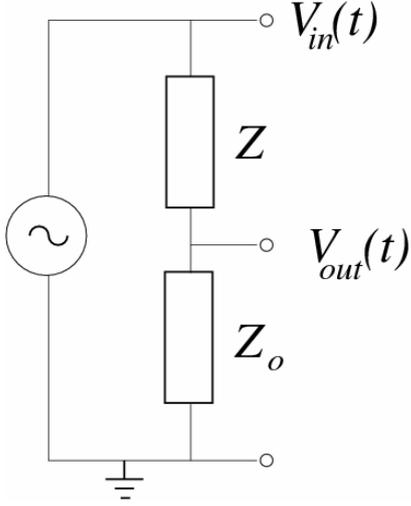

*Fig. 3: Scheme of the simple layout to measure the unknown impedance Z in terms of the value of a reference impedance $Z_0$.*

The unknown impedance Z is obtained in terms of the reference impedance $Z_0$ as

$$Z = R_Z + jX_Z = Z_0\left(\frac{|V_{in}|}{|V_{out}|}e^{-j\phi_0} - 1\right) = Z_0\left[\frac{\cos(\phi_0)}{A_0} - 1 - j\frac{\sin(\phi_0)}{A_0}\right] \tag{6}$$

where $\phi_0$ is the phase delay of the output signal with respect to the input and $A_0 = \frac{|V_{out}|}{|V_{in}|}$ is the output attenuation.

This method can be useful when measuring the reactive component of the unknown impedance but it suffers of two drawbacks that considerably reduce its accuracy in the measure of the resistance. Recalling the analysis of section II, we note that the experimental determination of the output attenuation $A_0$ depends on the values of the two overall calibration factors of the input, $K_{in}$, and output, $K_{out}$, channels, that are generally poorly known. Assuming as a guideline that these factors are equal to 1 with a relative uncertainty of 1/128 for our 7+1 bit ADC converters, we see that the attenuation $A_0$ is affected by a systematic uncertainty larger than 1% that can impact severely on the uncertainty affecting the value of $R_Z$.

The second drawback stems from the fact that, although we may try to compensate as well as we can the two probes used to pick-up signals, nevertheless the phase shift introduced by the two measuring channels of the oscilloscope are bound to differ by an amount that can severely affect our measurements.

Both these problems can be solved performing a new set of measurements of phase shift and attenuation with the same setup after replacing our unknown impedance Z with a known test impedance $Z_T$. We will obtain $Z_T = R_T + jX_T = Z_0\left(\dfrac{e^{-j\phi_T}}{A_T} - 1\right)$ that, combined with previous eq. (6) yields

$$Z = R_Z + jX_Z = (Z_0 + Z_T)\dfrac{A_T}{A_0}e^{j(\phi_T - \phi_0)} - Z_0 \qquad (7)$$

This formula shows that, when the two sets of measurements are performed with the same settings of the digital oscilloscope, the results for Z are independent from the voltage calibration factors and from any phase shift introduced by the measurement channels. In these conditions only statistical fluctuations and noise have to be taken into account for the attenuations yielding to the leading order

$$\dfrac{\sigma_A}{A} \cong \dfrac{1}{\sqrt{N}}\left[\left(\dfrac{\sigma_{in}^2 + <(\delta V_{noise}^{in})^2>}{<V_{in}(t_i)^2>}\right)^{1/2} + \left(\dfrac{\sigma_{out}^2 + <(\delta V_{noise}^{out})^2>}{<V_{out}(t_i)^2>}\right)^{1/2}\right]$$

Comparing with eq. (5) we see that the relative statistical uncertainty affecting attenuation is very similar to the absolute statistical phase uncertainty. With the use of eq. (7) their effect on the measured values of resistance and reactance can be easily evaluated.

Thin metal film resistors are the most appropriate choice for both $Z_0$ and $Z_T$ since they have lower tolerances and temperature coefficients, lower noise, lower parasitic inductance and capacitance. Moreover, their value can be measured to an accuracy of a few parts over thousands with low cost digital dc ohmmeter available in all student laboratories. The proper choice of their values depends on the reactance under test and the impact of their uncertainty on the measured value can be made comparable to the statistical uncertainties discussed above.

When aiming at measurements in the high frequency range, it is important to keep in mind that the impedance of the probe used to read the output signal affects the value of $Z_0$ that becomes then complex. When the knowledge of the probe capacity $C_p$ is only approximate, the value of $R_0$ should be sufficiently low to guarantee that the needed corrections are negligible. A similar consideration applies to the stray capacity $C_T$ across the test impedance $Z_T$. In this case an upper limit exists to the value of the resistance $R_T$ that in turns limits the useful frequency range for inductance measurements to $\omega << \frac{1}{\sqrt{LC_T}}$. As we will show in the following it is relatively easy, by choosing adequately the value of $R_0$, to maintain statistical uncertainties below 10% or 1% respectively for resistance or reactance measurements within this frequency limit.

Systematic effects due to trigger jitter are avoided by using the oscilloscope in single sweep option. The most serious parasitic effect affecting phase measurement is due to capacitive pickup and/or ground loop interference in the output channel. Therefore it is important that the layout of the experiment is carefully selected, keeping all conductor leads as short as possible, avoiding the use of jumper connections and using screened cables for signal routing.

It is always important to check the results we obtain against these possible interferences by working with multiple values of $R_0$ or $R_T$ for each sampled frequency. Also it is important to control that the results obtained are compatible within experimental uncertainties after exchanging the two measurements channel of the oscilloscope. However it should be noted that with these precautions we succeeded to perform the measurements illustrated below using a solder-less breadboard to connect components.

V  Resonant circuit experiments

In spite of its apparent simplicity, an experiment with a resonant RLC circuit leading to an accurate comparison with theoretical predictions requires a careful planning and an accurate choice of components and procedures. The main point to bear in mind is that the voltage across the two reactive components at resonance is higher than the input voltage by a factor equal to the quality factor Q of the filter. This can induce a current that is likely to be larger than the one used when testing the single components. To minimize the impact of this effect on the results a number of precautions are needed. First of all we should select an air core inductance for our experiment to avoid to be fooled by non linear properties of a magnetic core. Moreover, because of the higher dissipation caused by the higher current, we should use components with a low temperature sensitivity coefficient. This applies to the choice of the capacitor whose non linear dielectric properties are also a concern.

In the first experiment to be described here we used an air core inductor whose inductance L, as measured by a low frequency vectorial bridge, is equal to 2.790±0.005 mH, a polyester film capacitor whose capacity C, measured with the same bridge, is equal to 9.90±0.03 nF and a resistor whose resistance is 9.71±0.03 Ω. To minimize the residual thermal or non linear effects, we performed our experiments with an input signal of 100 mV to make sure not to exceed the voltage used during the following impedance measurements.

With these parameters we expect a resonant frequency of 30.373 kHz, in fair agreement with the experimental results, as shown in fig. 4. However we expect a quality factor in excess of 50 whereas a value of about 20 is measured. A similar discrepancy has been reported in published literature[2,5].

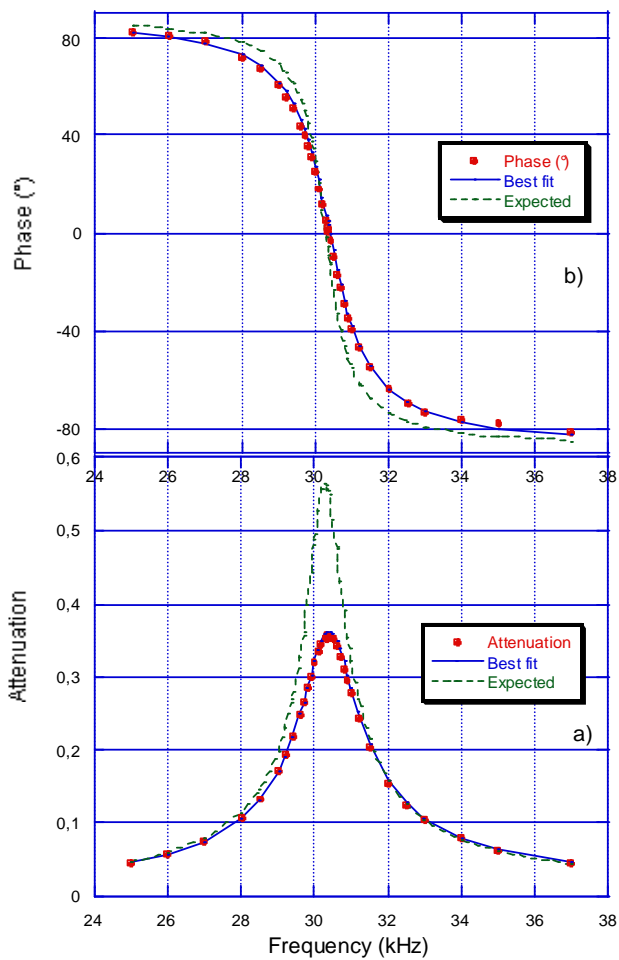

*Fig. 4: Frequency dependence of attenuation (a) and phase (b) of the voltage across the resistor R in the series RLC filter described in the text. The dashed line take into account only the dc resistance of the coil, the full line is a best fit to data.*

In the absence of dissipation in the reactive components we would also expect that the measured attenuation at the resonant frequency is equal to unity which is never the case in the experiment. Dissipation takes place in the inductor due to finite conductivity of its winding. Measuring its equivalent resistance $R_L$ with a dc ohmmeter may be adequate at low frequency but could prove wrong when the frequency is high enough to make the skin depth comparable to the wire radius. The dashed lines in fig. 5 take into account a measured dc value of 7.65 Ω for $R_L$ but still badly miss experimental data. As shown by the continuous lines in fig. 4, an excellent fit to data is obtained by arbitrarily increasing the total parasitic resistance to 17.2 Ω whose origin must be clarified.

Dissipation takes also place in the capacitor mainly due to dielectric losses which exhibit both intrinsic frequency dependence due dielectric behaviour and an extrinsic one due to reduction of the current leakage in the dielectric at increasing frequencies. With the laboratory vectorial bridge we can measure an equivalent series resistance $R_C$ strongly decreasing with frequency and reaching a value of 17 Ω at 10 kHz that is the high frequency limit of the instrument. Since this exceeds the amount needed to explain the data, we may try to extrapolate $R_C$ value at the resonant frequency taking into account only the extrinsic frequency dependence by assuming that the dielectric properties remain constant but this only yields a value of 2 Ω which is too low for our needs.

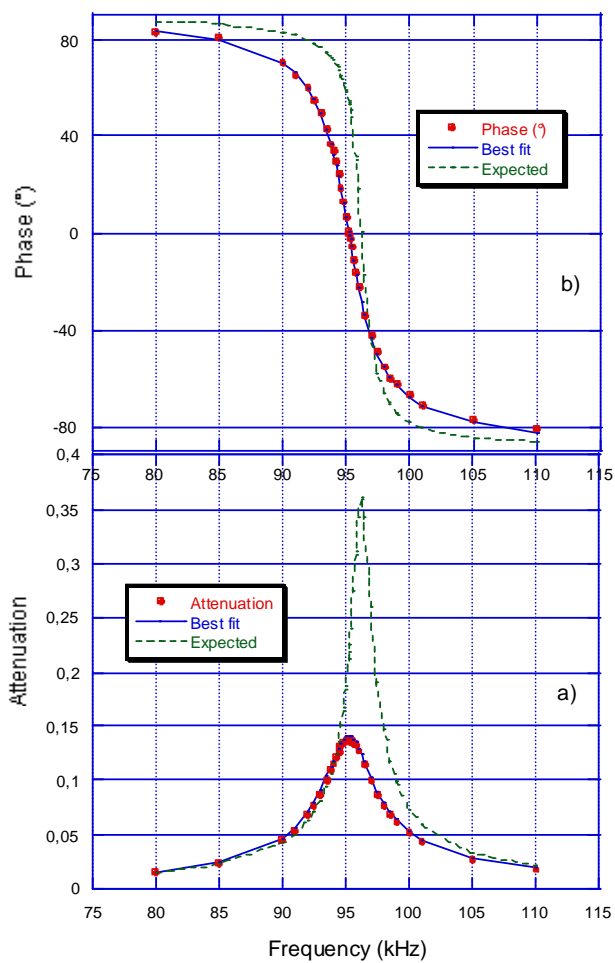

*Fig. 5: Frequency dependence of attenuation (a) and phase (b) of the voltage across the resistor R in the series RLC filter described in the text. The dashed lines use parameters of the low frequency case; the full lines are a best fit to data.*

Our second experiment has been conducted with the same components except for a new capacitor with capacity reduced by about a factor 10 (C = 0.980±0.003 nF) that we use in place of the previous one to increase the resonance frequency. The data obtained are displayed in fig. 5 again as full points. In this case the dashed lines in the figure are computed with the same parameters that fitted the previous experiment, except of course for the new capacity. The figure shows that we observe in this case a resonant frequency 1% lower than prediction. Now to fit the data we have to adjust the value of the inductance or the capacity and we need to increase the parasitic losses to an equivalent of about 60 Ω (see the continuous line in fig. 5). This frequency dependence of both losses and reactance remains to be explained.

## VI Capacitor and inductor impedance

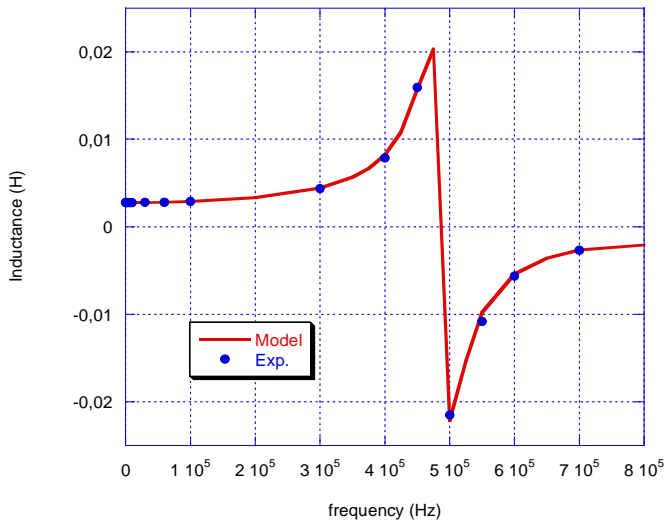

Fig. 6: Frequency dependence of inductance of the coil used in the series RLC filters. The line has been obtained with a lumped parameter model using the measured value of the auto-resonance frequency (488 kHz) and assuming a coil resistance of 500 Ω at the resonance.

To get a clearer answer to the questions arisen in the previous section, we carried out a detailed characterization of the components used in the experiments as a function of the excitation frequency. We adopted the method of section IV using a resistor for $Z_0$ whose resistance $R_0$, depending on the reactance of the component under test, could take one of the following three values: 9.72, 99.58 or 998.3 Ω. A resistor was also used for the test reactance $Z_T$ and its resistance $R_T$ was made comparable to the module of the unknown impedance at the sampled frequency. Both resistance and reactance were measured for each component and each frequency was sampled with at least two values of $R_T$. When multiple values were available we first checked that they were compatible with the respective accuracies and then we combined them in a weighted average. Moreover all data obtained at 1 kHz and 10 kHz were successfully checked against measurements performed with a dual frequency vectorial bridge available in the student laboratory. The values of the measured inductance for our air core inductor are displayed in fig. 6 in the frequency range 1 kHz to 700 kHz. This figure shows that L remains relatively constant up to 30 kHz and then starts to increase significantly. At 100 kHz we detect a relative increase of 3.5% with respect to the low frequency value that accounts for the resonance value in the second experiment discussed in the previous section. Negative values of the inductance at high frequency can only be explained by the presence of a stray inter-wire capacity. Adopting a lumped parameter model[13] to describe our coil and with the measured value of 488 kHz for the self resonance of our inductor, we get a fairly good quantitative account of the inductance dependence upon frequency. This is shown by the continuous line in the figure which represents the reactance of the coil in parallel with a stray capacity $C_s$ = 38 pF. The quality of the fit around the resonance region depends critically upon the resistance of the coil that needs to be increased to 500 Ω at the resonant frequency: this turns out to be consistent with the resistance measurements presented below.

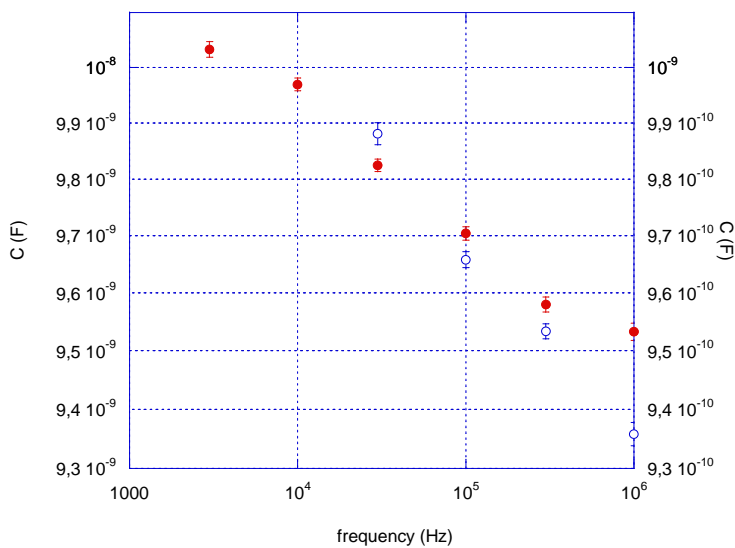

Fig. 7: Frequency dependence of the capacity of the two capacitors used in the series RLC filters. The right hand scale refers to the open points, while full dots refer to the left hand scale.

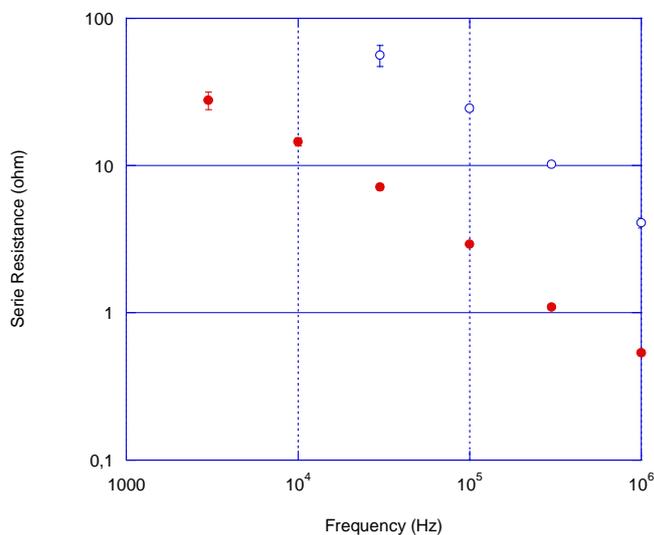

*Fig. 8: Frequency dependence of the series resistance of the two capacitors used in the series RLC filters. Full dots and open points correspond to 10 nF and 1 nF case respectively, see fig. 7.*

The capacity of the two capacitors used in the experiment turn out to decrease with frequency, see fig. 7. This reduction is similar for the two cases and is caused by the frequency dependence of the dielectric constant of the polyester used for isolation in both capacitors. Quantitatively it is in fair agreement with typical data provided by the manufacturer[14].

The equivalent series resistances of the two capacitors are shown in fig. 8. From these values we calculate a dissipation factor in agreement with manufacturer data. Apart for a scale factor, the frequency dependence is similar, as expected since the two components use the same dielectric.

When the value is higher than the resistance of leads and contacts, they scale inversely to their capacity, as justified from geometrical considerations.

The frequency dependence of the resistance of the inductor is much steeper, see fig. 9. An increase of 15% is observed at 30 kHz with respect to the dc value which becomes a factor 3.5 at 100 kHz where the skin depth becomes 0.2 mm, comparable with the wire radius of 0.15 mm. The peak observed at 500 kHz is caused by the parallel resonance with the stray inter-wire capacity discussed above.

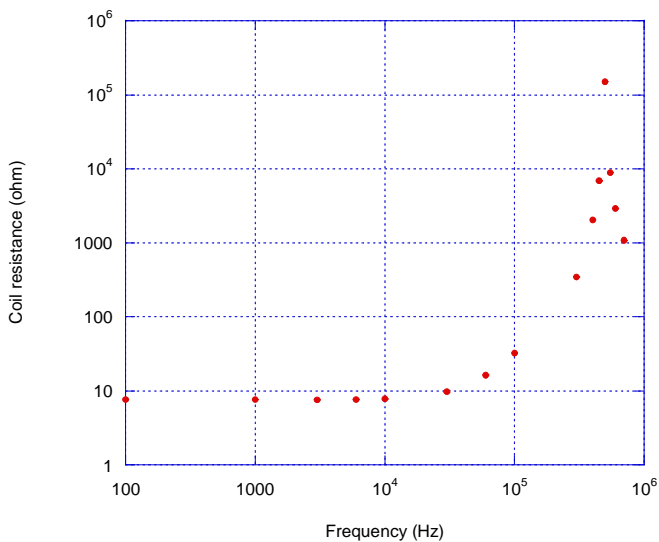

Fig. 9: Frequency dependence of resistance of the coil used in the series RLC filters.

The same model used to fit the reactance data can also be exploited to recover the naked resistance of the coil. These data are shown in fig. 10 where the resistance normalized to its low frequency value is plotted as a function of the frequency normalized to the value of 192.5 kHz yielding a current shin depth equal to the radius of the wire in the coil (0.15 mm). This figure shows that the parallel resonance is not the only cause of the increased coil resistance and justifies the assumption made in drawing fig. 6. It strongly indicates that eddy current effects become important well before the onset of the resonance.

In a wound coil two types of eddy currents are induced at high frequency[15]. Skin currents are responsible for enhanced losses in an isolated wire[16] or in a single layer coil: in this case the current is redistributed across the section of the wire and concentrated near to its surface. In a multilayer coil proximity effects are by far the most important: in this case in each layer the net current results from the difference of two currents running in opposite direction on the two faces of the layer. Since the magnitude of these currents increases with the number of layers, their effect can be felt when the skin depth is still larger then the wire dimension.

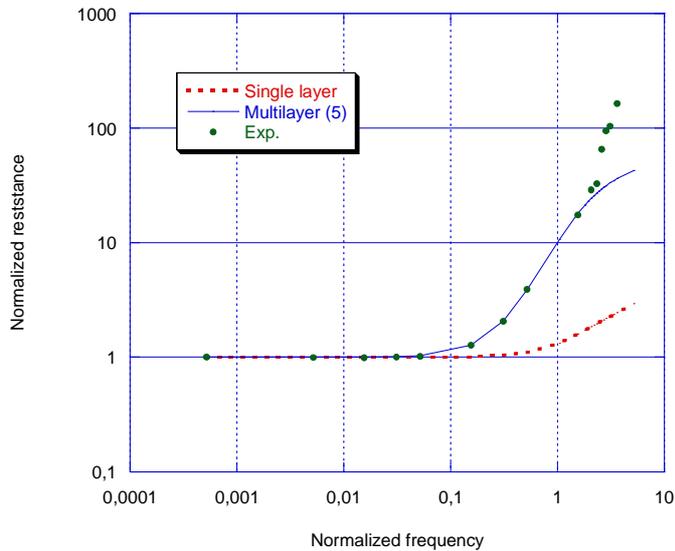

*Fig. 10: Frequency dependence of naked resistance of coil used in the series RLC filters. Dashed line accounts for skin effects only while continuous line includes proximity effects as well.*

Evaluation of the proximity effect can only be done numerically to get reasonable results at high frequency[17] and requires a detailed knowledge of the coil geometry which is not available to us. However we can obtain at least an approximate understanding of the importance of these effects using an analytic formulation[18]. In fig. 10 the dotted line represents the resistance enhancement caused by skin effect in a single layer (wire radius 0.15 mm) and is well below our experimental data. The full line takes into account the proximity effects and is computed for a 5 layer coil of the same wire diameter and a filling factor of 80%. We see that in this case the model has the capability to reproduce the experimental observations up to values of the skin depth lower than a few times the wire radius.

VII Conclusions

The data of the previous section allow for a better understanding of the observation reported in section V. They quantify the importance of additional loss mechanisms both in the capacitor, due to a frequency dependent dissipation caused by polarization currents in the dielectric, and in the inductor, where eddy currents and the parallel resonance with the inter-turn capacity greatly enhance the apparent resistance of the coil.

|  | LF | | HF | |
| --- | --- | --- | --- | --- |
|  | resonance | impedance | resonance | impedance |
| $\nu_0$ (Hz) | 30401±2 | 30330±65 | 95214±10 | 95024±185 |
| $R_L+R_C$ (Ω) | 17.19±0.02 | 16.9±0.5 | 59.7±0.1 | 57±1 |
| L (mH) | 2.790±0.007 | 2.803±0.005 | 2.91±0.02 | 2.905±0.004 |
| C (pF) | 9820±27 | 9820±11 | 958±6 | 966±2 |

Table II

In the narrow band-pass of our two filters we can neglect any frequency dependence and obtain an independent estimate of the inductance L, the capacity C and total losses $R_L+R_C$ at the resonant frequency $\nu_0$ by a fit of the resonance plots to their theoretical expectations. These values are reported, together with their uncertainties, in table II and compared with the corresponding measurements obtained with the vectorial bridge in the previous section. All the results are compatible within the experimental uncertainties indicating that the relevant physical effects have been taken into account.

In conclusion we have shown how the study of RLC circuits can be directed toward observations that can both motivate students to a critical reappraisal of models used to describe electrical components and introduce them to the physics of dielectric materials and electromagnetic effects in extended conductors.

A project that exploits the material presented in this paper could start with a first step consisting in the design, construction and characterization of a high Q band-pass filter. A discussion of results obtained should then lead to identify the necessity of a better evaluation of component impedances. The following step would consist in the theoretical study of phase measurements with digital oscilloscope giving special emphasis to the different source of uncertainties and their evaluation. In the third step the laboratory instruments should be reconfigured to build the vectorial bridge and impedance measurements and their uncertainties should be analysed. The frequency dependence of resistance and reactance of capacitors and inductors would then be the main subject of the forth step followed by a discussion of the relevant physics. A final step could then consist in the identification of better components to improve the quality of the filters, their construction and characterization.

Appendix

To evaluate the statistical uncertainties in the phase measurements caused by fluctuations in the value of the quantization errors it is useful to start by rewriting expression (3) as a function r of quantization errors

$$\cos(\phi) = r(\delta_1^{in}, \delta_2^{in} ..., \delta_N^{in}; \delta_1^{out}, \delta_2^{out} ..., \delta_N^{out}) = \frac{f(\delta_1^{in}, \delta_2^{in} ..., \delta_N^{in}; \delta_1^{out}, \delta_2^{out} ..., \delta_N^{out})}{g(\delta_1^{in}, \delta_2^{in} ..., \delta_N^{in}; \delta_1^{out}, \delta_2^{out} ..., \delta_N^{out})}$$

with

$$f(\delta_1^{in}, \delta_2^{in}..., \delta_N^{in}; \delta_1^{out}, \delta_2^{out}..., \delta_N^{out}) = <V_{in}(t_i)V_{out}(t_i)> - <V_{in}(t_i)><V_{out}(t_i)> =$$

$$= \frac{K_{in}K_{out}V_{in}V_{out}\cos(\phi)}{2} + \frac{1}{N}\sum_{i=1}^{N}[K_{in}V_{in}\sin(\omega t_i)(\delta_i^{out} + \delta_0^{out}) + K_{out}V_{out}\sin(\omega t_i + \phi)(\delta_i^{in} + \delta_0^{in})]$$

and

$$g(\delta_1^{in}, \delta_2^{in}..., \delta_N^{in}; \delta_1^{out}, \delta_2^{out}..., \delta_N^{out}) = \sqrt{<V_{in}(t_i)^2> - <V_{in}(t_i)>^2 - \sigma_{in}^2}\sqrt{<V_{out}(t_i)^2> - <V_{out}(t_i)>^2 - \sigma_{out}^2} =$$

$$= \sqrt{\frac{K_{in}^2 V_{in}^2}{2} + \frac{2}{N}\sum_{i=1}^{N}K_{in}V_{in}(\delta_i^{in} + \delta_0^{in})\sin(\omega t_i)}\sqrt{\frac{K_{out}^2 V_{out}^2}{2} + \frac{2}{N}\sum_{i=1}^{N}K_{out}V_{out}(\delta_i^{out} + \delta_0^{in})\sin(\omega t_i + \phi)}$$

Let us start with the contribution from the input channel. The sensitivity factors are given by

$\frac{\partial r}{\partial \delta_i^{in}} = \frac{1}{g^2}\left(g\frac{\partial f}{\partial \delta_i^{in}} - f\frac{\partial g}{\partial \delta_i^{in}}\right)$ where both $f$ and $g$, and their derivatives, are evaluated at the expected values $\delta_i^{in} = \delta_i^{out} = 0$. We have $\frac{\partial f}{\partial \delta_i^{in}} = \frac{K_{out}V_{out}\sin(\omega t_i + \phi)}{N}$ and $\frac{\partial g}{\partial \delta_i^{in}} = K_{out}V_{out}\frac{K_{in}V_{in}\sin(\omega t_i)}{N}$. Since the distribution of quantization errors can be assumed time independent, their contribution to the variance of $\cos(\phi)$ is given by

$$\sigma_{\cos(\phi)}^2 = \sigma_{in}^2\sum_{i=1}^{N}\left(\frac{\partial r}{\partial \delta_i^{in}}\right)^2 =$$

$$= \frac{\sigma_{in}^2}{N^2}\sum_{i=1}^{N}\left[\frac{K_{out}^2 V_{out}^2 \sin^2(\omega t_i + \phi)}{g^2} - \frac{f}{g}\frac{K_{out}^2 V_{out}^2 \sin(\omega t_i + \phi)\sin(\omega t_i)}{g^2} + \left(\frac{f}{g}\right)^2\frac{K_{out}^2 V_{out}^2 \sin^2(\omega t_i)}{g^2}\right] =$$

$$= \sigma_{in}^2 \frac{K_{out}^2 V_{out}^2}{Ng^2}\left(\frac{1}{2} - \cos^2(\phi) + \frac{\cos^2(\phi)}{2}\right) = \frac{2\sigma_{in}^2}{NK_{in}^2 V_{in}^2}\sin^2(\phi) = \frac{\sin^2(\phi)}{N}\frac{\sigma_{in}^2}{<V_{in}(t_i)^2>}$$

A similar expression can be recovered for the contribution of the output channel and, since the two channels are not correlated, expression (4) is now easily obtained.

As for the amplitude accuracy, starting from the expression

$$V_{in}^2 \equiv \langle V_{in}(t_i)^2\rangle = \frac{1}{N}\sum_{i=1}^{N}K_{in}^2 V_{in}^2 \sin(\omega t_i)^2 + 2K_{in}V_{in}(\delta_i^{in} + \delta_0^{in})\sin(\omega t_i) + (\delta_i^{in} + \delta_0^{in})^2$$

we can first compute the sensitivity factors as $\frac{2}{N} K_{in} V_{in} \sin(\omega t_i)$ and proceed as above to obtain

$$2V_{in}\sigma_{V_{in}} = \left[\sum_{i=1}^{N}\left(\frac{2K_{in}V_{in}}{N}\right)^2 \sin^2(\omega t_i)\sigma_{in}^2\right]^{\frac{1}{2}} = \left[\frac{2(K_{in}V_{in})^2}{N}\sigma_{in}^2\right]^{\frac{1}{2}}$$ which in turns yields

$$\frac{\sigma_{V_{in}}}{V_{in}} = \frac{1}{\sqrt{N}}\left[\frac{\sigma_{in}^2}{<V_{in}^2(t_i)>}\right]^{1/2}$$ from which the last equation in section III is recovered.